\def\pd #1#2{{\displaystyle{\partial {#1}} \over \displaystyle{\partial {#2}} }}
\def\ch{{\rm ch }\, }
\def\sh{{\rm sh }\, }
\def\aaj#1{{{\mathop {J}\limits_{#1}{}_{\!\!\mu\nu}} }}
\def\ax #1{{\mathop {X}\limits_{#1}}}
\def\ael #1{{\mathop {L}\limits_{#1}}}
\def\asum#1#2{\sum\limits_{#1} \limits^{#2}}

\documentclass[twoside,fleqn,11pt]{article}
\usepackage{latexsym,amsmath}      
                                                  
\begin{document}

\begin{flushleft}
\Large \bf 
Nonlinear Representations for Poincar\'e and
Galilei algebras and nonlinear equations
for electromagnetic fields
\end{flushleft}

\bigskip

\noindent
{\bf Wilhelm FUSHCHYCH~$^{\dag^1}$, Ivan TSYFRA~$^{\dag^2}$ 
and Vyacheslav BOYKO~$^{\dag^3}$}

\bigskip

\noindent
{\it Institute of Mathematics of the National Ukrainian Academy of Sciences,\\
3 Tereshchenkivska Str. 3, Kyiv 4, 01601 Ukraina}

\medskip

\noindent
$^{\dag^1}$~URL: {\tt http://www.imath.kiev.ua/\~{}appmath/wif.html}

\noindent
$^{\dag^2}$~E-mail: {\tt tsyfra@igph.kiev.ua}

\noindent
$^{\dag^3}$~URL: {\tt http://www.imath.kiev.ua/\~{}boyko/} \ \ 
E-mail: {\tt boyko@imath.kiev.ua}

\bigskip

\begin{abstract}
\noindent We construct nonlinear representations of the
Poincar\'e, Galilei, and conformal algebras on a set of the
vector-functions $\Psi =(\vec E, \vec H)$. A nonlinear complex
equation of Euler type for the electromagnetic field is proposed.
The invariance algebra of this equation is found.
\end{abstract}

\section{Introduction}

\noindent It is well known that the linear representations of the
Poincar\'e algebra $AP(1,3)$ and conformal algebra $AC(1,3)$, with
the basis elements
\begin{gather}
P_{\mu} =ig^{\mu \nu} \partial_{\nu}, \qquad
J_{\mu \nu}= x_{\mu} P_{\nu}-x_{\nu} P_{\mu} +S_{\mu\nu},
\\
       D=x_{\nu} P^{\nu} -2i,
\\
       K_{\mu}=2x_{\mu}D -(x_{\nu} x^{\nu})P_{\mu} +2x^{\nu}
       S _{\mu \nu},
\end{gather}
is realized on the set of solutions of the Maxwell
equations for the electromagnetic field in vacuum (see. e.g. \cite{1, 2})
\begin{gather}
\pd{\vec E}{t} ={\rm rot } \vec
H, \qquad {\rm div} \vec E = 0, \\
\pd{\vec H}{t} =-{\rm rot}
\vec E,\qquad {\rm div} \vec H = 0.
\end{gather}
Here $S_{\mu\nu}$ realize the representation $D(0,1)
\oplus D(1,0)$ of the Lorentz group.

Operators (1)--(3) satisfy the following commutation relations:
\begin{gather}
[P_{\mu},P_{\nu}] =0,\qquad [P_{\mu}, J_{\alpha \beta} ]=
    i(g_{\mu\alpha} P_{\beta}- g_{\mu\beta} P_{\alpha} ),\\
      [J_{\alpha \beta}, J_{\mu \nu} ]=i(g_{\beta \mu} J_{\alpha \nu} +g_{\alpha \nu} J_{\beta \mu}
                        -g_{\alpha \mu} J_{\beta \nu} -g_{\beta \nu} J_{\alpha \mu}),\\
         [D,P_{\mu}] = -i P_{\mu} ,\qquad [D, J_{\mu \nu}] = 0,\\
         [K_{\mu},P_{\alpha}] = i (2J_{\alpha \mu} -2g_{\mu \alpha} D),\qquad  [K_{\mu}, J_{\alpha \beta} ]
                           = i(g_{\mu \nu} K_{\beta} -g_{\mu \beta }K_{\alpha}),\\
         [K_{\mu},D ]= -i K_{\mu},\qquad [K_{\mu},K_{\nu}]=0,
                     \qquad \mu ,\,\nu ,\,\alpha ,\,\beta =0,\,1,\,2,\,3.
\end{gather}
In this paper the nonlinear representations of the
Poincar\'e, Galilei,  and conformal algebras for the
electromagnetic field $\vec E$, $\vec H$ are constructed. In
particular, we prove that the continuity equation for the
electromagnetic field is not invariant under the Lorentz group if
the velocity of the electromagnetic field is taken in accordance
with the Poynting definition. Conditional symmetry of the
continuity equation is studied. The complex Euler equation for the
electromagnetic field is introduced. The symmetry of this equation
is investigated.

\section{Formulation of the main results}

The operators, realizing the nonlinear representations of the
Poincar\'e
algebras 
$AP(1,3)=\langle P_{\mu}, J_{\mu \nu} \rangle$, $AP_1
(1,3)=\langle P_{\mu}, J_{\mu \nu},D \rangle $,  and conformal
algebra $AC(1,3)=\langle P_{\mu}, J_{\mu \nu},D
,K_{\mu}\rangle$,  have the structure
\begin{gather}
P_{\mu} =\partial_{x_{\mu}},\\
            J_{kl}=x_k \partial_{x_l} -x_l \partial_{x_k} +S_{kl},\\
J_{0k}=x_0 \partial_{x_k} +x_k \partial_{x_0} +S_{0k},\qquad
k,l=1, 2, 3,\\
                          D=x_{\mu} \partial_{x_{\mu}},\\
K_0=x_0^2 \partial_{x_0} +x_0 x_k \partial_{x_k} +(x_k
-x_0E^k) \partial_{E^k} -x_0H^k\partial_{H^k},\\ K_l=
x_0x_l\partial_{x_0} +x_lx_k\partial_{x_k} +[x_k E^l -x_0(E^l
E^k -H^l H^k )] \partial_{E^k}+ {}\nonumber\\
\qquad {}+ [x_kH^l -x_0 (H^l E^k +E^l H^k )] \partial_{H^k},
\end{gather}
where
\begin{gather*}
S_{kl} = E^k \partial_{E^l} -E^l \partial_{E^k} +H^k
\partial_{H^l}-H^l \partial_{H^k} ,\\ S_{0k} = \partial_{E^k} - (E^k
E^l - H^k H^l ) \partial_{E^l} - (E^k H^l + H^k E^l )
\partial_{H^l}.
\end{gather*}
The operators, realizing the nonlinear representations
of the Galilei algebras  $AG^{(2)} (1,3) =\langle
P_{\mu}, J_{kl},G_k^{(2)} \rangle ,$ $AG_1^{(2)} (1,3) =\langle
P_{\mu}, J_{kl},G_k^{(2)},D \rangle $ have the form:
\begin{gather}
 P_{\mu} =\partial_{x_{\mu}},\qquad J_{kl} =
x_k\partial_{x_l} -x_l\partial_{x_k} +S_{kl},\\
G^2_k =x_k \partial_{x_0} -(E^k E^l -H^k H^l ) \partial_{E^l} -
(E^k H^l +H^k E^l) \partial_{H^l},
\\
D= x_{0} \partial_{x_{0}}+2x_{k} \partial_{x_{k}}
+E^k\partial_{E^k}+H^k\partial_{H^k}.
\end{gather}

We see by direct verification that all represented operators
satisfy the commutation relations of the algebras $AP(1,3)$,
$AC(1,3)$, $AG(1,3)$. 

\section{Construction of nonlinear
representations}

In order to construct the nonlinear representations of
Euclid-, Poincare-, and Galilei groups and their extensions the
following idea was proposed in \cite{2,3}: to use nonlinear equations
invariant under these groups; it is necessary to find (point out,
guess) the equations, which admit symmetry operators having a
nonlinear structure. Such equation for the scalar field
$u(x_0,x_1,x_2,x_3) $ is the eikonal equation
\begin{gather}
\pd{u}{x_{\mu}} \pd{u}{x^{\mu}} =0,\qquad \mu=0, 1, 2, 3 
\end{gather}
which is invariant under the conformal algebra $AC(1,3)$
with the nonlinear operator $K_{\mu}$ \cite{2, 3}.

The nonlinear Euler equation for an ideal fluid
\begin{gather}
\frac{\partial v_k}{\partial t} +
v_l \frac{\partial v_k}{\partial x_l} =0,\qquad k=1,
2, 3 
\end{gather}
which is invariant under nonlinear representation of the
$AP(1,3)$ algebra, with basis elements
\begin{gather}
P_{\mu} =\partial_{x_{\mu}},\qquad
J_{kl} =x_k \partial_{x_l} -x_l \partial_{x_k}
     + v_k \partial_{v_l} - v_l \partial_{v_k}, \\
J_{0k} =x_k \partial_0 +x_0 \partial_{x_k} +\partial_{v_k}
-v_k v_l \partial_{v_l},
\end{gather}
was proposed in \cite{3} to construct the nonlinear
representation for the vector field. Note that equation (21) is
also invariant with respect to the Galilei algebra $AG(1,3)$ with
the basis elements
\begin{gather}
P_{\mu} =\partial_{x_{\mu}},\qquad J_{kl} =
x_{k} \partial_{x_l} -x_l \partial_{x_k}
              + v_k \partial_{v_l} - v_l \partial_{v_k}, \qquad
               G_a =x_0 \partial_{x_a} +\partial_{v_a}.
\end{gather}
As mentioned in \cite{2, 3}  both the
Lorentz--Poincar\'e--Einstein and Galilean principles of
relativity are valid for  system (21). We use the following
nonlinear system of equations~\cite{4}
\begin{gather}
\pd {E^k}{x_0} + H^l \pd {E^k} {x_l}=0,
\qquad \pd {H^k} {x_0} + E^l\pd {H^k} {x_l}=0,
\end{gather}
for constructing a nonlinear representation of the
$AP(1,3)$ and $AG(1,3)$ algebras for the electromagnetic field. To
construct the basis elements of the $AP(1,3 )$ and $AG(1,3)$
algebras in explicit form we investigate the symmetry of system
(25). We search for the symmetry operators of equations (25) in
the form:
\begin{gather}
X= \xi^{\mu} \partial_{x_{\mu}} +\eta^l
\partial_{E^l} + \beta^l \partial_{H^l},
\end{gather}
where $\xi^{\mu} =\xi^{\mu} (x, \vec E, \vec H )$,
$\eta^l =\eta^l (x, \vec E, \vec H )$, $\beta^l=\beta^l (x, \vec E,
\vec H ).$

\medskip

\noindent {\bf Theorem 1.} {\it The maximal invariance algebra
of system (25) in the class of ope\-ra\-tors (26) is the
20-dimensional algebra, whose basis elements are given by the
formulas}
\begin{gather}
P_{\mu}=\partial{x_{\mu}},\\
J^{(1)}_{kl}= x_k \partial_{x_l} -x_l \partial_{x_k} +E^k
\partial_{E^l}-E^l \partial_{E^k}
                                             +H^k\partial_{H^l} -H^{l}\partial_{H^k},\\
J^{(2)}_{kl}= x_k \partial_{x_l} +x_l \partial_{x_k} +E^k
\partial_{E^l}+E^l \partial_{E^k}
                                             +H^k\partial_{H^l} +H^{l}\partial_{H^k},\\
G_a^{(1)}=x_0 \partial_{x_a} +\partial_{E^a}
+\partial_{H^a},\\
G_a^{(2)}=x_a \partial_{x_0} -E^a E^k
\partial_{E^k} - H^a H^k \partial_{H^k},\\ 
D_0=x_0
\partial_{x_0} -E^l  \partial_{E^l} - H^l  \partial_{H^l},\\
D_1=x_1 \partial_{x_1} +E^1  \partial_{E^1} + H^1 \partial_{H^1},\\
 D_2=x_2 \partial_{x_2} +E^2  \partial_{E^2}+ H^2  \partial_{H^2},\\ 
D_3=x_3 \partial_{x_3} +E^3
\partial_{E^3} + H^3  \partial_{H^3}.
\end{gather}

\noindent {\bf Proof.} To prove theorem 1 we use Lie's
algorithm. The condition of invariance of the system $L(\vec E,
\vec H)$, i.e. (25), with respect to operator $X$ has the form
\begin{gather}
\ax 1 \ael {}\Bigl|_{L=0} =0,
\end{gather}
where
\begin{gather*} \ax 1 =\ax {}+[ D_{\alpha} (\eta^l) -E^l_j D_{\alpha}
(\xi^j)] \partial_{E^l_{\alpha}} +[D_{\alpha} (\beta^l) -H_j^l D_{\alpha}
(\xi^j)]\partial_{H^l_{\alpha}},\\
 E^l_{\alpha} = \pd {E^l} {x_{\alpha}},\quad H^l_{\alpha} = \pd{H^l}{x_{\alpha}},\qquad l=1, 2, 3;
                                     \quad \alpha =0, 1, 2, 3
\end{gather*}
is the prolonged operator. From the invariance condition
(36) we obtain  the system of equations which determine the
coefficient functions $\xi^{\mu}$, $\eta^l$, $\beta^l$ of the operator
(26): 
\begin{gather}
 \eta^l_k =0,\qquad\eta_0^l=0,\nonumber\\
\beta^l_k =0,\qquad\beta_0^l=0,\qquad \xi_{\alpha \nu}^{\mu}=0,\nonumber\\
\xi_{E^a} ^{\mu} =0,\qquad \xi_{H^a}^{\mu}=0,\nonumber\\ \eta^k
=-E^k \xi^0_0 +\xi ^k_0 +E^a \xi_a^k -E^a E^k \xi^0_a,\\
\beta^k =-H^k \xi^0_0 +\xi ^k_0 +H^a \xi_a^k -H^a H^k
\xi^0_a,\nonumber
\end{gather}
where
\begin{gather*}
\eta^l_k =\pd {\eta^l} {x_{k}},\qquad
 \eta^l_0 =\pd{\eta^l} {x_{0}},\qquad
 \xi^{\mu}_{E^a}=\pd {\xi^{\mu}} {E^a},
\qquad
 \xi^{\mu}_{\alpha \nu}={\partial^2 \xi^{\mu} \over \partial
x_{\alpha} \partial x_{\nu} } .
\end{gather*}
Having found the general solution of system (37), we get
the explicit form of all the linear independent symmetry operators
of system (25),  which have the structure  (27)--(35). 
Operators of Lorentz rotations $J_{0k}$ is given by the linear
combination of the Galilean operators $G_k^{(1)}$ and $G_k^{(2)}$:
\begin{gather}
J_{0k}=G_k^{(1)} +G_k^{(2)}.
\end{gather}
All the following statements, given here without proofs,
can be proved in analogy with the above-mentioned scheme.

\section{The finite transformations and
invariants}

We present some finite transformations which are
generated by the operators $J_{0k}$:
\begin{align}
 J_{01}: \quad &  x_0\to x_0'= x_0 \ch \theta_1 +x_1\sh
\theta_1,\nonumber\\
    & x_1\to  x_1'= x_1 \ch \theta_1 +x_0\sh \theta_1,\\
      &  x_2 \to x'_2 =x_2,\qquad x_3 \to x'_3 =x_3,\nonumber\\
        & E^1 \to E^{1'} ={ E^1 \ch \theta_1 +\sh \theta_1
\over E^1 \sh \theta_1 +\ch
\theta_1},\qquad
        H^1 \to H^{1'} ={ H^1 \ch \theta_1 +\sh \theta_1
\over H^1 \sh \theta_1 +\ch\theta_1},\nonumber\\
        & E^2 \to E^{2'} ={ E^2\over E^1 \sh\theta_1 +\ch\theta_1},\qquad
        H^2 \to H^{2'} ={ H^2\over H^1 \sh \theta_1 +\ch
\theta_1},\\
    &     E^3 \to E^{3'} ={ E^3\over E^1 \sh \theta_1 +\ch\theta_1},\qquad
        H^3 \to H^{3'} ={ H^3\over H^1 \sh\theta_1 +\ch\theta_1}.\nonumber
\end{align}
The operators $J_{02}$, $J_{03}$ generate analogous
transformations. $\theta_1$ is the real group parameter of the
geometric Lorentz transformation. Operators $G_k^{(2)}$ generate
the following transformations:
\begin{align*}
G_{1}^{(2)}:\quad &  x_0\to x'_0 = x_0 + \theta_1 x_1
,\qquad
       x_k \to x'_k =x_k ,\\
&        E^k \to E^{k'} ={E^k \over 1 + \theta_1 E^1},\qquad
        H^k \to H^{k'} ={H^k \over 1 + \theta_1 H^1}.
\end{align*}
Analogous transformations are generated by the operators
$G_2^{(2)}$, $G_3^{(2)}$. Operators $G_k^{(1)}$ generate the
following transformations:
\begin{align*}
G_{1}^{(1)}:\quad & x_0\to x'_0 =
x_0 ,\qquad x_1 \to x'_1 =x_1 +x_0 \theta_1,\\
&           x_2\to x'_2 = x_2 ,\qquad  x_3 \to x'_3 =x_3,\\
&            E^1 \to E^{1'} =E^1 +\theta_1,\qquad
           H^1 \to H^{1'} =H^1 +\theta_1,\\
&           E^2 \to E^{2'} =E^2,\qquad E^3 \to E^{3'} =E^3,\\
 &          H^2 \to H^{2'} =H^2,\qquad H^3 \to H^{3'} =H^3.
\end{align*}
The operators $G_2^{(1)}$, $G_3^{(1)}$ generate analogous
transformations.

It is easy to verify that
\begin{gather}
I_1 ={\left(1- \vec E \vec H\right)^2 \over \left(1-\vec E^2 \right)
\left(1-\vec H^2\right)}, \qquad \vec E^2 \ne
1,\qquad \vec H^2 \ne 1
\end{gather}
is invariant with respect to the nonlinear
transformations of the Poincar\'e group which are generated by
representations (28), (38).

The invariant of the Galilei group which is generated by
representations (28), (31) has the form:
\begin{gather}
I_2 ={ \vec E^2  \vec H^2  \over
\left( \vec E  \vec H\right)^2 },
\end{gather}
 whereas the Galilei group which is generated by
representations (28), (30) has the invariant
\begin{gather}
 I_3 =( \vec E - \vec H)^2.
\end{gather}

\section{Complex Euler equation for the
electromagnetic field}

Let us consider the system of equations
\begin{gather}
{\partial \Sigma^k \over \partial x_0} +
\Sigma^l   {\partial \Sigma^k \over \partial x_l}=0,\qquad
                       \Sigma^k =E^k +iH^k.
\end{gather}
The complex system (44) is equivalent to the real system
of equations for $\vec E$ and $\vec H$
\begin{gather}
{\partial E^k \over \partial x_0 }
+ E^l  {\partial E^k \over \partial x_l } - H^l
{\partial H^k \over \partial x_l } =0,\nonumber\\
{\partial H^k \over \partial x_0 } +  H^l{\partial
E^k \over \partial x_l } + E^l{\partial H^k
\over \partial x_l } =0.
\end{gather}
The following statement has been proved with the help of
Lie's algorithm.

\medskip

\noindent {\bf Theorem 2.} {\it The maximal invariance algebra
of the system (45) is the 24-dimen\-si\-o\-nal Lie algebra whose
basis elements are given by the formulas}
\begin{gather}
 P_{\mu} =\partial_{x_{\mu}},\nonumber\\
J^{(1)}_{kl} =x_k\partial_{x_l} -x_l\partial_{x_k} +E^k
\partial_{E^l}-E^l \partial_{E^k}
                               +H^k \partial_{H^l}-H^l \partial_{H^k},\nonumber\\
J^{(2)}_{kl} =x_k\partial_{x_l} +x_l\partial_{x_k} +E^k
\partial_{E^l}+E^l \partial_{E^k}
                                 +H^k \partial_{H^l}+H^l \partial_{H^k},\nonumber\\
G_a^{(1)} =x_0 \partial_{x_a} +\partial_{E^a},\nonumber\\
G_a^{(2)}= x_a \partial_{x_0} - (E^a E^k -H^aH^k) \partial_{E^a}
                                - (E^a H^k +H^aE^k) \partial_{H^k}, \\
D_0=x_0 \partial_{x_0} - E^k \partial_{E^k} -H^k
\partial_{H^k},\nonumber\\
 D_a=x_a \partial_{x_a} + E^a \partial_{E^a}
+H^a \partial_{H^a} \ \ \ \ \ {\mbox{ (no sum over}}\  a),\nonumber\\
K_0=x_0^2 \partial_{x_0} +x_0 x_k \partial_{x_k} + (x_k -x_0
E^k) \partial_{E^k} -x_0 H^k \partial_{H^k},\nonumber\\ K_a=x_0 x_a
\partial_{x_0} +x_a x_k \partial_{x_k} + [x_kE^a  -x_0(E^a E^k
-H^a H^k)]\partial_{E^k}+{}\nonumber\\
\qquad {}+ [ x_k H^a -x_0 (H^a E^k
+E^a H^k)] \partial_{H^k}. \nonumber
\end{gather}
 The algebra, engendered by the operators
(46), include the Galilei algebras $AG^{(1)} (1,3)$, 
$AG^{(2)} (1,3)$ and Poincar\'e algebra $AP(1,3)$, and conformal
algebra $AC(1,3)$ as subalgebras. Operators $G^{(2)}_a$ generate
the linear geometrical transformations in $R(1,3)$
\begin{gather}
x_0 \to x_0' =x_0 +\theta_a x_a \ \ \ \ \ {\mbox
{(no sum over}}\  a),\qquad x_l \to x'_l,
\end{gather}
as well as the nonlinear transformations of the fields
\begin{gather}
 E^l + iH^l \to E^{l'}+i H^{l'} = {E^l +iH^l \over 1+ \theta_a (E^a +iH^a) } 
\qquad {\mbox {(no sum over}}\  a), \nonumber\\
 E^l- iH^l \to E^{l'} - i H^{l'} = {E^l -iH^l
\over 1+ \theta_a (E^a -i H^a)}.
\end{gather}

The invariant of the group $G^{(2)} (1,3)$ is
\begin{gather}
I_4 = { \bigl( \vec E^2 -\vec H^2\bigr) + 4 \bigl(\vec E \vec H\bigr)^2
\over \bigl( \vec E^2 +\vec H^2 \bigr)^2 }.
\end{gather}
Operators $J_{0k} $ generate the linear transformations
in $R(1,3)$
\begin{gather} x_0 \to x'_0 =x_0 \ch \theta_k +x_0 \sh
\theta_k,\nonumber\\
 x_k \to x'_k =x_k \ch \theta_k +x_0 \sh
\theta_k\qquad {\mbox { (no sum over}}\  k ),\\
x_l\to x'_l =x_l, \qquad \mbox{if} \  l\ne k,\nonumber
\end{gather}
as well as the nonlinear transformations of the fields
\begin{gather*}
E^k +i H^k \to E^{k'} +i H^{k'} =  {
(E^k +iH^k) \ch \theta_k +\sh \theta_k
\over   (E^k +iH^k) \sh \theta_k +\ch
\theta_k },
\\
   E^k -i H^k \to E^{k'} -i H^{k'} =
 { (E^k -iH^k) \ch \theta_k +\sh
\theta_k \over (E^k -iH^k) \sh \theta_k
+\ch \theta_k }.
\end{gather*}
If  $l\ne k$, then 
\begin{gather}
E^l +i H^l \to E^{l'} +i H^{l'} = 
{E^l +iH^l \over (E^k +iH^k) \sh
\theta_k +\ch \theta_k},\nonumber\\
                                                         E^l -i H^l \to E^{l'} -i H^{l'} = {E^l -iH^l
\over (E^k -iH^k) \sh \theta_k +\ch
\theta_k} 
                        \qquad {\mbox {(no sum over}}\  k).
\end{gather}
The invariant of group $P(1,3)$ is
\begin{gather}
I_5={ 1-2 \left[ (\vec E^2 -\vec
H^2)-{1\over 2} (\vec E^2 -\vec H^2)^2
-2 (\vec E\vec H)^2\right] \over \left[1
-(\vec E^2 +\vec H^2) \right]^2 }, \qquad \vec E^2
+\vec H^2\ne 1. 
\end{gather}
The operator $K_0$ generates the following  nonlinear
transformations in $R(1,3)$ and linear transformations of the
fields
\begin{gather}
x_{\mu} \to x'_{\mu} ={x_{\mu}
\over 1- \theta_0 x_0},\nonumber\\
 E^k \to E^{k'} =E^k +\theta_0 (x_k -x_0 E^k),\\
 H^k \to H^{k'} =H^k (1 -\theta_0 x_0).\nonumber
\end{gather}
The operators $K_a$ generate nonlinear transformations
in both $R(1,3)$ and of the fields
\begin{gather*}
x_0 \to x'_0 ={x_0 \over 1-x_a \theta_a},\qquad x_a \to x'_a ={x_a \over 1-x_a\theta_a}.
\end{gather*}
If $k\ne a$, then
\begin{gather*}
x_k \to x'_k ={x_k \over 1-x_a\theta_a},\\
 E^a +iH^a \to E^{a'} +i H^{a'}
={E^a +i H^a \over 1 +\theta_a [x_0
( E^a +i H^a ) -x_a ]} ,\\
 E^a -iH^a \to E^{a'} -i
H^{a'} ={E^a -i H^a \over 1+\theta_a [x_0 ( E^a -i H^a ) -x_a ]}.
\end{gather*}
If $k\ne a$, then
\begin{gather}
E^k +iH^k \to E^{k'} +i H^{k'} = {E^k +i H^k +\theta_a (E^a +i H^a)
x_k \over 1 +\theta_a [x_0 (E^a +i H^a
) -x_a ]} ,\\
 E^k -iH^k \to E^{k'} -i H^{k'} ={E^k -i H^k +
\theta_a (E^a -i H^a) x_k \over 1 +\theta_a [x_0 (E^a -i H^a ) -x_a ]} \qquad
{\mbox{(no sum over}}\  a).\nonumber
\end{gather}

\noindent {\bf Note 1.} Setting $\asum {}{\to} =a \vec E +i b
\vec H$, where $a$, $b$ are arbitrary functions of the invariants
$\vec E^2$, $\vec H^2$, $\vec E \vec H$, we obtain more general
form of the equation (44). The equation
\[
{\partial \Sigma^k \over \partial x_0} +
\Sigma^l     {\partial \Sigma^k \over \partial x_l} =F
(\vec E \vec H, \vec E^2, \vec H^2 ) \Sigma^k 
\]
is invariant only under some subalgebras of algebra 
(46) depending on the choice of function~$F$.

\medskip

\noindent {\bf Note 2.} If we analyse the symmetry of the
following equations
\begin{gather}
 \left({ \partial \over \partial x_0} + E^l
{ \partial \over \partial x_l} + H^l { \partial \over \partial
x_l} \right) E^k=0,\nonumber\\
   \left({ \partial \over \partial x_0} +
E^l { \partial \over \partial x_l} + H^l { \partial \over
\partial x_l} \right) H^k=0;\tag{*}
\end{gather}
or
\begin{gather}
{ \partial E^k \over \partial x_0}
=\pm \left( E^l { \partial \over \partial x_l} + H^l {\partial \over \partial x_l } \right) H^k,\nonumber\\
{\partial H^k \over \partial x_0} =\pm \left( E^l {\partial \over \partial x_l} + H^l { \partial \over \partial x_l}
\right) E^k,\tag{**}
\end{gather}
we obtain  concrete examples of nonlinear
representations for the Poincar\'e and Galilei algebras. This
problem will be considered in a future paper. 

\section{Symmetry  of  the continuity   equation\\
and the Poynting vector}

Let us consider the continuity equation for the
electromagnetic field
\begin{gather}
L (\vec E, \vec H) \equiv {\partial \rho \over \partial x_0} +{\rm div } \rho \vec v
=0.
\end{gather}
According to the Poynting definition $\rho$ and $\rho
v^k$ have the forms
\begin{gather}
\rho ={1 \over 2}  (\vec E^2 + \vec
H^2), \qquad \rho v^k =\varepsilon_{k ln} E^l H^n.
\end{gather}

\noindent {\bf Theorem 3.} {\it The nonlinear system (55),
(56) is not invariant under the Lorentz algebra, with  basis
elements:}
\begin{gather}
J_{kl} =x_k \partial_{x_l}  -x_l
\partial_{x_k} +E^k \partial_{E^l}- E^l \partial_{E^k} +H^k \partial_{H^l}
-H^l \partial_{H^k},\nonumber\\
 J_{0k} =x_k \partial_{x_0}  +x_0 \partial_{x_k} +\varepsilon_{kln} (E^l \partial_{H^n} -H^l \partial_{E^n}),
\qquad k, l, n=1, 2, 3.
\end{gather}

To  prove theorem 3 it is necessary to substitute $\rho$ and $\rho
v^k$, from formulas (56), to  equation (55) and to apply Lie's
algorithm, i.e., it is necessary to verify that the invariance
condition
\begin{gather}
\aaj{1} \left( L (\vec E,\vec H )\right) \Bigr|_{L=0} \equiv 0
\end{gather}
is not satisfied, where $\aaj{1} $ is the first
prolongation of the operator $J_{\mu \nu}$.

\medskip

\noindent {\bf Theorem 4.} {\it The continuity equation (55),
(56) is conditionally invariant with respect to the operators
$J_{\mu \nu}$, given in  (57) if and only if $\vec E$, $\vec H$
satisfy the Maxwell equation (4), (5).}

\medskip

Thus the continuity equation, which is the  mathematical
expression of the conservation law of the electromagnetic field
energy and impulse is not Lorentz-invariant if $\vec E$, $\vec H$
does not  satisfy the Maxwell equation. A more detailed discussion
on conditional symmetris can be found in \cite{1,2}.

The following statement can be proved in the case when
\begin{gather}
\rho = F^0 (\vec E, \vec H)\qquad  \mbox {and}\qquad
\rho v^k =F^k (\vec E, \vec H ),
\end{gather}
where $F^0$, $F^k$ are arbitrary smooth functions $F^0
\not\equiv 0$, $F^k \not \equiv 0$. 

\medskip

\noindent {\bf Theorem 5.} {\it The continuity equation (55),
(59) is invariant with respect to the classic geometrical Lorentz
transformatons if and only if
\begin{gather}
r(B) = 4, 
\end{gather}
where $r(B)$ is the rank of the Jacobi matrix of
functions $F^{\mu}$. }

\medskip

In conclusion we present some statements about the symmetry of the
following systems:
\begin{gather}
{\partial \vec E \over \partial x_0}= {\rm rot}\, \vec H +\vec F_1 (\vec E, \vec
H),\qquad 
{\partial \vec H \over \partial x_0}= -{\rm rot}\, \vec E +\vec F_2 (\vec
E, \vec H),\nonumber\\
{\rm div}\, \vec E  =R_1 (\vec E, \vec H),
\qquad {\rm div}\, \vec H = R_2 (\vec E, \vec
H),
\end{gather}
\begin{gather}
{\partial (R\vec E) \over \partial
x_0}={\rm rot}\, (R\vec H),\qquad
{\partial N\vec H \over \partial x_0}= -{\rm rot}\,(N \vec E),\nonumber\\
{\rm div}\, (R \vec E)= 0,\qquad {\rm div}\,(N \vec H)= 0.
\end{gather}
Here
\begin{gather*}
 R=R(W_1,W_2),\qquad N=N(W_1,W_2),\\
 W_1=\vec E^2 -\vec H^2,\qquad W_2=\vec E \vec H.
\end{gather*}
\begin{gather}
{\rm div} (R\vec E +N\vec H)=0.
\end{gather}

\noindent {\bf Theorem 6.} {\it The  system of equations (61)
is invariant under the Lorentz algebra with the basis elements
(57) if and only if}
\[
\vec F_1 \equiv\vec F_2 \equiv 0, \qquad
R_1\equiv R_2   \equiv 0.
\]

\noindent {\bf Theorem 7.} {\it The  system of equations (62)
is invariant under the Lorentz algebra (57) if $R$ and $N$ are
arbitrary functions of the invariants $W_1 =\vec E^2 -\vec H^2$,
$W_2 =\vec E \vec H$. }

\medskip

\noindent {\bf Theorem 8.} {\it The equation (63) is invariant
under the Lorentz algebra with the basis elements (57) if and only
if $\vec E$, $\vec H$ satisfy the system of equations}
\[
{ \partial (R \vec E + N \vec H)
\over \partial x_0 }= {\rm rot}\, (R\vec H -N\vec E).
\]

Thus it is established that, besides the generally
recognized linear representation of the Lorentz group discovered
by Henry Poincare in 1905 \cite{5}, there exists the nonlinear
representation constructed by using the nonlinear equations of
hydrodynamical type~\cite{4}. It is obvious that for instance  the
linear superposition principle does not hold for a non-Maxwell
electrodynamic theory
 based on the equation (25) or (45).

The nonlinear representations for the algebras $AG(1,3)$, $A\tilde
P (1,2)$, $A\tilde P (2,2)$, $AC(1,2)$, $AC(2,2)$ for a scalar
field have been considered in~\cite{6}, $AP(1,1)$ in~\cite{7}, and $AP(1,2)$
in~\cite{8}.

\end{document}